\documentclass[twocolumn,aps,a4paper,showpacs,showkeys,preprintnumbers,amsmath,amssymb]{revtex4}
\usepackage{isolatin1}
\newcommand{\beq}{\begin{equation}}
\newcommand{\eeq}{\end{equation}}
\newcommand{\beqw}{\begin{widetext}\begin{equation}}
\newcommand{\eeqw}{\end{equation}\end{widetext}}
\newcommand{\beqa}{\begin{eqnarray}}
\newcommand{\eeqa}{\end{eqnarray}}
\renewcommand{\(}{\left(}
\renewcommand{\)}{\right)}
\newcommand{\R}{\mathbb{R}}
\newcommand{\Z}{\mathbb{Z}}
\newcommand{\G}{\mathcal{G}}
\newcommand{\K}{\mathcal{K}}
\newcommand{\LL}{\mathcal{L}}
\newcommand{\T}{\mathcal{T}}
\newcommand{\1}{\mathbf{1}}
\newcommand{\F}{\mathbf{F}}
\renewcommand{\c}{\mathbf{c}}
\newcommand{\f}{\mathbf{f}}
\newcommand{\n}{\mathbf{n}}
\renewcommand{\r}{\mathbf{r}}
\renewcommand{\t}{\mathbf{t}}
\renewcommand{\d}{\mbox{d}}
\newcommand{\ddx}{\frac{\partial}{\partial x}}
\newtheorem{The}{Theorem}
\newtheorem{Def}{Definition}
\newtheorem{Rem}{Remark}
\begin{document}
%%%%%%%%%%%%%%%%%%%%%%
\title{A General Formalism for Inhomogeneous Random Graphs}
\date{July 19, 2002}
\author{Bo Söderberg}
\email{Bo.Soderberg@thep.lu.se}
%\homepage{http://www.thep.lu.se/complex/}
\affiliation{Complex Systems Division, Dept. of Theoretical Physics,
Lund University, Sölvegatan 14A, S-223 63 Lund}
\begin{abstract}
We present and investigate an extension of the classical random graph
to a general class of inhomogeneous random graph models, where
vertices come in different types, and the probability of realizing an
edge depends on the types of its terminal vertices. This approach
provides a general framework for the analysis of a large class of
models. The generic phase structure is derived using generating
function techniques, and relations to other classes of models are
pointed out.
\end{abstract}
\preprint{LU TP 02-29 (to appear in {\bf Phys. Rev. E})}
\maketitle
%%%%%%%%%%%%%%%%%%%%%%

%%%%%%%%%%%%%%%%%%%%%%
\section{Introduction}
%%%%%%%%%%%%%%%%%%%%%%

The concept of random graphs (RG) has recently become the target of an
increasing interest, as a tool for modeling various kinds of networks,
arising e.g. in physics, biology and biophysics as well as in social
and information-technological structures.

The classical RG model \cite{ErRe60,Boll85,Jans00} describes a
homogeneous, sparse random graph of order $N$, where each edge is
randomly and independently realized with a fixed probability
$p=c/N$. For large orders, there is a critical value of $c = 1$, above
which almost every graph contains a single giant connected component
being of order $O(N)$, with the remaining components being small
compared to $N$. This model yields an asymptotic degree distribution
that is Poissonian with the average degree given by $c$.

Many real-life networks, such as the internet, have been shown to
possess other types of degree distribution, sometimes displaying a
power law behavior over many orders of magnitude, ruling out the
classical RG as the relevant model. A number of alternative RG models
have been suggested in an attempt to yield random graphs with more
general types of degree distribution, such as the desired power
behavior.
Some of these models describe dynamical random graphs, where the
graphs arise as the result of a stochastic growth process, such as
randomly grown networks \cite{Call01,Turo02}, or scale-free networks
based on preferential attachment\cite{AlBa00}. Others focus on describing
ensembles of random graphs with certain given properties, without
bothering about how they came about; a particularly interesting
approach of this type, possessing a high degree of generality, is
based on considering random graphs of fixed order with a given
arbitrary degree distribution \cite{Lucz92,MoRe95,MoRe98,Newm01}.

In this article, we will investigate a general class of sparse {\em
inhomogeneous} RG models, by means of a straightforward generalization
of the classical model to a situation where vertices may come in
different {\em types}, such that the probability for an edge depends
on the types of its pair of terminal vertices. While this class of
models inherits certain features from the homogeneous model -- such as
the existence of a critical hypersurface in parameter space, beyond
which asymptotically almost every graph has a giant component --, it
is capable of producing a wide class of asymptotic degree
distributions, among these distributions with power law behavior.

This general class of models is shown to contain a number of existing
models as special cases, and can be used as a general framework for
the analysis of various RG models.

The structure of this article is as follows. In Sec. 2, some of the
more salient features of the classical model are briefly reviewed,
while our generalization is presented and analysed in Sec. 3. In
Sec. 4, a number of special cases are discussed, while Sec. 5 contains
our conclusions.

%%%%%%%%%%%%%%%%%%%%%%%%%%%%%%%%%%%%%%%%%%%%%
\section{The Classical Model}
%%%%%%%%%%%%%%%%%%%%%%%%%%%%%%%%%%%%%%%%%%%%%

Here will briefly review some of the more prominent properties of the
classical random graph in the large $N$ limit, to pave the ground for
the subsequent analysis of its generalization.
\begin{Def}
Let $\G(N,c)$, with $c$ a real positive number, denote the
ensemble of graphs of order $N>c$, where each edge is independently
realized with probability $p = c/N$.
\end{Def}
This ensemble has a critical value of $c=1$, above which almost all
graphs for large $N$ have a single large connected component -- the
giant component -- with a finite fraction of the vertices, while the
remaining components are small.

%%%%%%%%%%%%%%%%%%%%%%%%%%%%%%%%%%%%%%%%%%
\subsection{Exposing Connected Components}

The standard method to reveal the size distribution of the orders of
components is to expose these as follows. Start with a single (random)
vertex, reveal its neighbors by following edges, then their neighbors,
etc. Let $n_k$ be the number of vertices exposed for the first time in
step $k$ of this process. The distribution of $n_k$, given the
previous numbers $n_0=1,n_1,\dots n_{k-1}$, becomes
\beq
	P(n_k) = \binom{N - \sum_{l=0}^{k-1} n_l}{n_k}
	\( 1 - q^{n_{k-1}} \)^{n_k}
	\( q^{n_{k-1}} \)^{N-\sum_{l=0}^k n_l},
\eeq
where $q=1-c/N$.

In the large $N$ limit with a fixed $c$, $P(n_k)$ tends to
$e^{-n_{k-1}c}(n_{k-1}c)^{n_k}/n_k!$, and the process reduces to a
Poissonian branching tree model $\mathcal{B}(c)$, with each vertex
independently branching to a number of new vertices, where this number
is a Poissonian random variable with average $c$.
The distribution $p_n$ over the order $n$ of the resulting tree is
conveniently analyzed in terms of the generating function $F(z)=\sum_n
p_n z^n$, which must satisfy
\beq
\label{F}
	F(z) = z \exp(c (F(z) - 1)).
\eeq
This can be solved iteratively for each $z$, and $F(z)$ must be a
stable fixed point of the corresponding iterated map; it is easy to
see that this implies $|F(z)| < 1 / |c|$. For $|z| \leq 1$, there is a
unique solution for $F(z)$, reachable from the starting point 0, given
by $F(z) = \hat{C}(z)/c$, where $\hat{C}(z)$ is the unique solution to
$xe^{-x} = zce^{-c}$ in the unit disk. Expanding the
corresponding inverse of $xe^{-x}$ yields the exact result $p_n =
n^{n-1} c^{n-1} e^{-nc}/ n!$.

Particularly interesting is the result for $z=1$, defining $f\equiv
F(1)$, which represents the total probability and might be expected to
be 1, which is an obvious solution to Eq. (\ref{F}) for $z=1$. Indeed,
for $c<1$, this solution is stable, but for $c>1$, it becomes
unstable, and another fixed point becomes the attractor, given by $f =
\hat{c}/c$, where $\hat{c}$ is the unique solution to
$\hat{c}e^{-\hat{c}} = ce^{-c}$ in the interval $[0,1]$.

Thus, for $c<1$ the branching model is subcritical, and always
terminates after a finite number of steps, while for $c>1$ it is
supercritical -- the deficit in total probability is due to a finite
probability $1-f$ that the order $n$ of the generated tree becomes
infinite, i.e. that the branching process never terminates.

For a large but finite $N$, this corresponds to all components being
small, i.e. $o(N)$, for $c<1$, while for $c>1$ there exists a single
giant component of order $\sim N(1-f)$ with the remaining components
being small, having an order distribution similar to that obtained for
the complementary $c$-value $\hat{c}$.

%%%%%%%%%%%%%%%%%%%%%%%%%%%%%%%%%%%%%%%%%%%%%%%
\section{Generalization to Inhomogeneous Random Graphs}
%%%%%%%%%%%%%%%%%%%%%%%%%%%%%%%%%%%%%%%%%%%%%%%

The classical RG model can be generalized in a straight-forward way to
inhomogeneous graphs by assuming that vertices can come in different
{\em types} $i \in \{1,\dots,K\}$. This enables us to consider a very
general class of inhomogeneous RG models, to be referred to as {\em IRG}:
\begin{Def}
Given a positive integer $K$, a $K$-dimensional vector
$\r=\{r_1\dots r_K\}$ of positive probabilities summing to 1, and a
symmetric $K\times K$ matrix $\c$ with non-negative elements $c_{ij}$,
let $\G(N, K, \r, \c)$ denote the ensemble of graphs $G$ of order $N$,
defined as follows:
\\ i) Each vertex is independently assigned a type $i\in \{1\dots K\}$
with probability $r_i$.
\\ ii) Independently for each unordered pair of vertices, the
corresponding undirected edge is realized with probability $p_{ij} =
c_{ij}/N$, where $(i,j)$ is the corresponding pair of vertex types.
\end{Def}
\begin{Rem}
An asymptotically equivalent alternative is to fix the number of
vertices of each type to certain values $N_i \approx N r_i$, and
possibly also the number of edges between vertices of types $i,j$ to
fixed values $E_{ij} \approx \(1-\delta_{ij}/2\) c_{ij}N_iN_j/N$.
\end{Rem}

%%%%%%%%%%%%%%%%%%%%%%%%%%%%%%%%%%%%%%%%%%%%%%%%
\subsection{Revelation of a Connected Component}

In analogy to the classical $\G(N,c)$ model, the model $\G(N, K, \r,
\c)$ can be analyzed by recursively revealing a connected component by
exploring neighbors, starting from a single vertex.
Let $n_{i,k}$ be the number of new vertices of type $i$ revealed in
the $k$th stage of the revelation (so $n_{i,0}=\delta_{i,i_0}$, with
$i_0$ the type of the starting vertex). Given the number of revealed
vertices of different types in the previous stages, $n_{i,k}$ obeys
the conditional distribution
\beqw
\label{Pnik}
	P(n_{i,k}) = \binom{N_i - \sum_{l=0}^{k-1} n_{i,l}}{n_{i,k}}
	\( 1 - \prod_j q_{ij}^{n_{j,k-1}} \)^{n_{i,k}}
	\( \prod_j q_{ij}^{n_{j,k-1}} \)^{N_i-\sum_{l=0}^k n_{i,l}},
\eeqw
where $q_{ij} = 1 - p_{ij} = 1 - c_{ij}/N \approx \exp(-c_{ij}/N)$.
This expression can be simplified in different domains.

%%%%%%%%%%%%%%%%%%%%%%%%%%%%%%%%%%%%%%%%%%%%%%%%%%%%%%%%%%%%%%%%%%%%%%%%%%%%%%%
\subsection{Small Components and the Branching Process Approximation}

As long as the order of the revealed part is small as compared to $N$,
we can approximate Eq. (\ref{Pnik}) by the Poisson distribution
\beq
\label{Pnik_s}
	P(n_{i,k}) \approx
	e^{-\sum_j r_i c_{ij} n_{j,k-1}}
	\frac{\(\sum_j r_i c_{ij} n_{j,k-1}\)^{n_{i,k}}}
	{n_{i,k}!}.
\eeq
This corresponds to approximating the revelation process by a
Poissonian random branching process.

For the distribution $P(n)$ of the total number $n$ of generated
vertices stating from a random vertex, we can define a {\em generating
function} $F(z)=\sum_n P(n) z^n$.  Since the distribution will depend
on the type of the initial vertex, $F$ must be written as the weighted
average of the corresponding generating functions $F_i(z)$ for the
distributions $P_i(n)$ conditional on the initial type $i$, i.e.
\beq
	F(z) = \sum_i r_i F_i(z), \;\;\;\mbox{with}\;\;\;
	F_i(z) = \sum_n P_i(n) z^n.
\eeq
The vector $\F$ having the different $F_i$ as components satisfies the
coupled set of equations
\beq
\label{Fi}
	F_i(z) = z \exp\(\sum_j c_{ij} r_j \(F_j(z) - 1\)\),
\eeq
which is the inhomogeneous version of Eq. (\ref{F}).
\begin{Rem}
A more elaborate set of generating functions $\tilde{F}_i(z_1\dots
z_K) = \sum_{\n} P_i(\n) \prod_j z_j^{n_j}$ could be defined, using a
distinct variable $z_i$ for each type $i$, with $n_i$ the total number
of revealed vertices of type $i$. These would obey equations obtained
by replacing the ``$z$'' after the equal sign in Eq. (\ref{Fi}) by
``$z_i$''. Here we do not care about the detailed type content, and
the simpler version, $F_i(z) = \tilde{F}_i(z,\dots,z)$, will suffice.
\end{Rem}
Interpreting Eq. (\ref{Fi}) as a $K$-dimensional iterated map (replace
``$=$'' by ``$:=$''), the proper solution is the stable fixed point
reached from the starting point $\F=0$.
%
% In what follows we will restrict the discussion to $F_i(z) \equiv
% F_i(z,z\dots,z)$, yielding the generating function for the
% distribution $P_i(n)$ over the order $n=\sum_k n_k$ independently of
% type content, but conditional on the type of the inital vertex,
% $F_i(z)=\sum_n P_i(n) z^n$.
%
Particularly interesting is the result for $z=1$, so let $f_i =
F_i(1)$, expressing the probability that the branching process will
terminate, conditional on the type of the starting vertex, and let the
unconditional counterpart be denoted by $f=\sum_j r_j f_j=F(1)$. The
$f_i$ satisfy the coupled set of equations
\beq
\label{fi}
	f_i = \exp\(\sum_j c_{ij} r_j \(f_j - 1\)\),
\eeq
with a naive solution $\f=\1$, the stability of which can be analyzed
by means of linearization of Eq. (\ref{fi}) around $\f=\1$, yielding
$\{c_{ij}r_j\}$ as the relevant matrix.
This is all we need in order to pin down the appearance of the giant,
as well as its asymptotic size, and we state the result without proof
(it follows by analogy to the corresponding result for the classical
model):
%
%We are now ready to formulate one of the main results of this article,
%which we state without proof (it should be provable in analogy to the
%corresponding result for the classical RG model):
%
\begin{The}
A) The model $\G(N,K,\r,\c)$ is subcritical if the eigenvalues of
the matrix $\{c_{ij}r_j\}$ are all less than one in absolute value;
the graphs then a.a.s. possess no giant component.\\B) When some
eigenvalue is larger than one, the model is supercritical, and the
graphs a.a.s. possess a giant component; its number $n_i$ of vertices
of type $i$ asymptotically satisfies $n_i/N \sim r_i(1-f_i)$, where
the $f_i$ correspond to a stable solution of Eq. (\ref{fi}).
\end{The}
\begin{Rem}
It appears natural to require in addition that $\c$ cannot be
block-diagonalized; otherwise ergodicity would be broken, and the
graph would trivially decompose into distinct subgraphs, which could
be treated separately.
\end{Rem}
In the supercritical case, the generating functions $F_i(z)$ can be
{\em renormalized} with $f_i$, to yield generating functions for the
finite (non-giant) component part. Let $\hat{F}_i(z)=F_i(z)/f_i$. Then
$\hat{\F}$ is a stable solution of
\beq
   \hat{F}_i(z) = z \exp\( \sum_j c_{ij} r_j f_j (\hat{F}_j - 1)\),
\eeq
with $\hat{\F}(1) = \1$.  This describes a subcritical branching
process with renormalized parameters $\hat{r}_i = r_i f_i / \r\cdot\f$
and $\hat{c}_{ij} = c_{ij} \;\r\cdot\f$.  For a finite $N$, we must have
$\hat{N} \sim N\;\r\cdot\f$, and we see that this conserves
$\hat{p}_{ij} = \hat{c}_{ij}/\hat{N} = p_{ij}$; thus, the renormalized
model is simply the naive restriction of the original one to the
subset of vertices outside the giant component.

%%%%%%%%%%%%%%%%%%%%%%%%%%%%%%%%%%%%%%%%%%%%%%%%%%%%%%%%%%%%%%%%%%%%%%%%%%%%%%%%
\subsection{Large Components and the Deterministic Approximation}

When the giant component is revealed, another approximation can be
made to Eq. (\ref{Pnik}). Once the number of revealed vertices become
of $O(N)$, the distribution of $n_{i,k}$ becomes sharply peaked around
its average, due to a factor of $N$ appearing in the exponent. As a
result, the fluctuations become negligible, yielding a deterministic
iterative equation for the consecutive revealed numbers. In terms of
the fraction $g_{i,k} = 1 - \sum_{l=0}^k n_{i,l}/(Nr_i)$ of all
vertices of type $i$ not yet revealed after step $k$, this yields
\beq
\label{det2}
	g_{i,k} =
	g_{i,k-1} \exp\(\sum_j c_{ij} r_j (g_{j,k-1} - g_{j,k-2})\),
\eeq
revealing the conserved quantities
\beq
	\mu_i \equiv g_{i,k} \exp\(\sum_j c_{ij} r_j \(1-g_{j,k-1}\)\).
\eeq
The values of $\mu_i$ must be $\sim 1$, since their values can only
change in an earlier stage when the number of revealed vertices is
still small, but then $g\sim 1$; thus, in the large $N$ limit we can
safely assume $\mu_i = 1$. The two-step recursion (\ref{det2}) reduces
to a one-step recursion, taking the form $g_{i,k} = e^{\sum_j c_{ij}
r_j \(g_{j,k-1}-1\)}$, which can be seen as iterating the map
\beq
\label{gi}
	g_i \to \exp\(\sum_j c_{ij} r_j \(g_j - 1\)\)
\eeq
until a stable fixed point is reached. If the model is subcritical,
this is given by the trivial fixed point $g_i=1$, whereas for a
supercritical model a non-trivial fixed point with $g_i<1$ results,
signalling the existence of a giant component containing a fraction
$1 - g_i$ of the vertices of type $i$.

Eq. (\ref{gi}) is identical to Eq. (\ref{fi}), which was derived in
the limit of small numbers of revealed vertices; thus, we have
established the same set of equations in two different limits.
\begin{Rem}
A third, heuristic way of estimating the size of the giant component
is as follows. Suppose the giant contains a fraction $n_i$ of the
vertices of type $i$. Then we can estimate its neighborhood, i.e. the
set of vertices connected to at least one vertex in the giant (which
of course must be the giant itself), as follows, based on the rather
bold assumption that the edge probabilities do not depend on whether
any or both of its terminal vertices are in the giant: The total
number of vertices of type $i$ is $~ N r_i$. For each of these, the
probability of {\em not} being connected to any of the vertices in the
giant is $\exp(-\sum_j c_{ij} n_j/N)$. Thus we can expect a number $N
r_i \(1 - \exp(-\sum_j c_{ij} n_j/N)\)$ of vertices of type $i$ in the
neighborhood, i.e. in the giant. Writing $n_i$ as $N r_i (1-g_i)$, we
recover Eq. (\ref{gi}), in spite of the bold assumptions involved.
\end{Rem}

%%%%%%%%%%%%%%%%%%%%%%%%%%%%%%%%%%%%%
\subsection{Extended Type Spaces}

While we have assumed a finite number of types $K$, defining the {\em
type space} $\T=\Z_K$, the above results should be more or less
directly extendable to models where the type space $\T$ is a
denumerable infinite set, or even a continuous manifold, under some
general conditions yet to be precisely determined.
\begin{Def}
For a given type space $\T$, with a normalized measure $r$ on
$\T$, and a given non-negative symmetric function $c(x,y)$ on $\T^2$,
define $\G(N,\T,\r,\c)$ as the ensemble of RGs of order $N$, where
each vertex is independently assigned a type $x\in\T$ according to
$r(x)$, and for each vertex pair the corresponding edge is
independently chosen with probability $c(x,y)/N$, with $(x,y)$ the
corresponding pair of types.
\end{Def}
For the {\em denumerable} case, $\T=\Z_+$, it appears natural to
require the asymptotic degree averages $c_{ij}$, or at least the total
averages $C_i=\sum_{j=1}^{\infty} c_{ij} r_j$, to be uniformly
bounded.  For cases where the elements of $\c$ are unbounded, an
alternative is to regularize $p_{ij}$ for finite $N$ by using $p_{ij}
= 1 - \exp(-c_{ij}/N)$ instead of the unbounded $c_{ij}/N$.

Also, reasonable care may have to be taken that $\c$ is sufficiently
ergodic. Let $t_{ij}$ be 0 if $c_{ij}=0$, 1 otherwise. The matrix
$\t$ then describes a graph in type space, with $t_{ij}=1$
corresponding to the existence of the edge $(i,j)$. Then, sufficient
ergodicity could e.g. mean that this graph should have a finite
diameter, i.e. a uniformly bounded distance between vertex pairs.

For the case of a {\em continuous} type space $\T$, similar care must
be taken. In addition, some kind of continuity constraint seems
appropriate, both on $\c$ and $\r$.

Note that a continuous $\T$ allows for a continuous {\em
reparameterization invariance}. Thus, for the case of $\T=\R$, assume
$f$ to be a strictly increasing, continuously differentiable mapping
of $\R$ to itself. Then the model defined by
$\hat{c}(x,y)=c(f(x),f(y))$ and $\hat{r}(x) = r(f(x)) f'(x)$ is
completely equivalent to the one with $c(x,y)$ and $r(x)$.  Thus,
$r(x)$ could be transformed to any desirable normalized distribution
on $\R$. In particular, it could be transformed to the uniform
distribution on the unit interval, yielding a kind of standard
representation of the model. For higher-dimensional manifolds, things
are more complicated, and it appears difficult to devise a universal
standardization procedure.

A precise determination of feasibility conditions for extended type
spaces will be the subject of future work.

%%%%%%%%%%%%%%%%%%%%%%%%%%%%%
\subsection{Degree Distributions}

Many properties (but not all!) of a graph ensemble are reflected in
its asymptotic degree distribution.  In IRG, the asymptotic degree
distribution $p_m$ is determined by $\r$ and $\c$, and given simply as
the weighted average of the type-specific degree distributions
$p_{m|i}$, being Poissonian with an average $C_i$ defined by $C_i =
\sum_j c_{ij} r_j$. The result is
\beq
	p_m = \sum_i r_i \exp\(-C_i\) \frac{C_i^m}{m!}
\eeq
with the associated generating function
\beq
\label{H}
	H(z) \equiv \sum_m p_m z^m = \sum_i r_i \exp\(C_i (z - 1)\).
\eeq

This puts a limitation on the possible degree distributions that can
be obtained within IRG: It must be possible to write the distribution
as a positive linear combination of Poissonians, i.e.
\beq
	p_m = \frac{1}{m!} \int_0^{\infty} c^m e^{-c} p(c) \d c,
\eeq
where $p(c)\geq 0$ describes an, {\em a priori} arbitrary,
distribution of type-specific Poissonian degree averages $c=C_i$,
assuming the possibility of a continuum of types.
This is a kind of smoothness constraint. In particular, it implies
\beq
	p_m^2 \leq \frac{m+1}{m} p_{m+1} p_{m-1}
\eeq
for each $m>0$. While this excludes e.g. random regular graphs where
the degree is fixed, it does allow for a wide class of degree
distributions, such as distributions with a power tail, $p_m\propto
m^{-\alpha}$ for large $m$, by letting $p(c)$ having a similar power
tail, $p(c)\propto c^{-\alpha}$ for large $c$.

Note that a particular model in IRG is {\em not} determined solely by
the degree sequence, which depends on ${c_{ij}}$ only through the
average $C_i=\sum_jc_{ij}r_j$. This is in contrast to a class of
recently considered models \cite{Lucz92,MoRe95,MoRe98,Newm01}; such
models define a particular subclass of IRG, however, as will be shown
below.

%%%%%%%%%%%%%%%%%%%%%%%%%%%%%%%%%%%
\section{Special Cases of Interest}
%%%%%%%%%%%%%%%%%%%%%%%%%%%%%%%%%%%

For $K=1$, of course the known properties of the classical RG model is
recovered. Below we will consider a few less trivial examples.

%%%%%%%%%%%%%%%%%%%%%%%%%%%%%%%%%%%
\subsection{Random Bipartite Graph}

Assuming two distinct vertex types, i.e. $K=2$, a simple ensemble of
random bipartite graphs results from the choice
\beq
	\c = \(
	\begin{array}{cc}
	0 & a \\ a & 0
	\end{array}
	\).
\eeq
With an arbitrary choice of type distribution $\r=(r_1,r_2)$, this
yields for the asymptotic generating function $\F(z) = (F_1,F_2)$ the
equations
\begin{subequations}
\beqa
	F_1(z) &=& z \exp\(a r_2 \(F_2(z) - 1\)\),
\\
	F_2(z) &=& z \exp\(a r_1 \(F_1(z) - 1\)\).
\eeqa
\end{subequations}
For $z=1$, this yields
\beq
  f_1 = \exp\(a r_2 \(f_2 - 1\)\), \qquad f_2 = \exp\(a r_1 \(f_1 - 1\)\),
\eeq
yielding the critical value of $a$ as $a_c = 1/\sqrt{r_1 r_2}$.
In the symmetric case of $r_1=r_2=1/2$, we have $a_c=2$, and
$f_1=f_2=f$ satisfying $f = \exp\(a/2 \(f - 1\)\)$.

In a similar way, ensembles of random $K$-partite graphs can be
defined, which can be seen as generated by the complete graph
$\K_K$. Similarly, ensembles of random graphs based on an arbitrary
generating graph can be defined, with $\c$ proportional to the
incidence matrix for the generating graph.  A nice twist results from
using a random graph as a generator.

%%%%%%%%%%%%%%%%%%%%%%%%%%%%%%%%%%%%%%%%%%%
\subsection{Rank-1 $\c$ Matrix}

A particularly interesting special case results when $\c$ has the
factorized form $c_{ij} = C_i C_j/\bar{C}$, where $C_i > 0$ can be
interpreted as a connection tendency for vertices of type $i$, while
$\bar{C}=\sum_i r_i C_i$.

Writing the asymptotic generating function as $F(z)=\sum_i r_i
F_i(z)$, we get for $F_i(z)$ in this case
\beq
   F_i(z) = z \exp\(\frac{C_i \sum_j r_j C_j \(F_j(z) - 1\)}{\bar{C}}\),
\eeq
which can be reduced to a single equation for the function $G(z) =
\sum_i r_i C_i F_i(z)/\bar{C}$, reading
\beq
	G(z) = z \sum_i r_i C_i \exp\( C_i (G(z) - 1) \) / \bar{C}.
\eeq
%
% H(z) = \sum_i r_i \exp\(C_i (z - 1)\)
% H'(z) = \sum_i r_i C_i \exp\(C_i (z - 1)\)
% H'(1) = \sum_i r_i C_i = \bar{C}
% H''(z) = \sum_i r_i C_i^2\exp\(C_i (z - 1)\)
% H''(1) = \sum_i r_i C_i^2
In terms of the generating function $H(z)$ for the asymptotic degree
distribution, Eq. (\ref{H}), this can be written as
\beq
\label{Gz}
	G(z) = z \frac{H^{'}(G(z))}{H^{'}(1)},
\eeq
%
% F_i(z) = z \exp\( C_i (G(z) - 1) \)
and in terms of $G(z)$ we have
\beq
\label{Fz}
	F(z) = z H(G(z)).
\eeq
For $z=1$ in particular, we get for $g=G(1)$ the equation
\beq
	g = \frac{H^{'}(g)}{H^{'}(1)},
%g' = H''/H'
\eeq
and linearization around the trivial solution $g=1$ yields stability
for $H^{''}(1)/H^{'}(1) < 1$, corresponding to the model being {\em
subcritical} for $\langle C^2 \rangle < \langle C \rangle$, which is
equivalent to $\langle m^2 \rangle < \langle 2m \rangle$ in terms of
moments of the degree distribution.

%--------------------------------------%
%\subsubsection{Relation to Other Models}

With $\c$ restricted to have rank 1, the resulting models are
asymptotically equivalent to models from a superficially very
different class of random graphs that has recently attracted some
attention \cite{Lucz92,MoRe95,MoRe98,Newm01}. There, a random graph
ensemble based on an arbitrary asymptotic degree distribution $p_m$ is
defined for a finite order $N$ by randomly selecting a member from the
set of graphs with a given degree sequence, such that the number of
vertices with degree $m$ is approximately $Np_m$.

Also for such a model, the recursive exposition of a connected
component asymptotically yields a well-defined branching process,
apparently very different from the Poissonian ones obtained for IRG.
Here, the inital vertex is assigned a random degree $m$ according to
$p_m$, and subsequently branches to $m$ daughter vertices. Each new
vertex is independently assigned a degree $n>0$, distributed according
to $n p_n/\sum_m m p_m$ (consistent with the assumption that the
asymptotic probability of connecting to a particular vertex is
proportional to its degree), and then branches to $n-1$ daughters
(since one of its edges is already used).

The asymptotic generating function $F(z) = \sum_k P_k z^k$ for
the resulting order distribution $P_k$ then satisfies the equation
\beq
	F(z) = z \sum_m p_m G(z)^m,
\eeq
expressing the choice of the initial degree $m$. Here, $G(z)$ is the
{\em edge generating function}, which satisfies
\beq
	G(z) = z \frac{\sum_m m p_m G(z)^{m-1}}{\sum_m m p_m},
\eeq
expressing the choice of the daughter's degree $m$, and its branching
to $m-1$ edges.

These are nothing but Eqs. (\ref{Fz}) and (\ref{Gz}) in disguise,
showing the complete asymptotic equivalence of the two models, despite
the superficial differences; indeed, the criteria $\langle m(m-2)
\rangle<0$ for subcriticality derived above are in complete accordance
with the results of Ref. \cite{MoRe95}.

%%%%%%%%%%%%%%%%%%%%%%%%%%%%%%%%%
\subsection{Dynamical Random Graph with Finite Memory}

The last example is given by a recently proposed class of dynamical
random graphs \cite{Turo02} with memory, where a graph is produced
starting from a single node according to the combination of three
random processes in continuous time, all Poissonian:
\begin{enumerate}
\item For each existing vertex, new, initially isolated, vertices are
      added at a rate $\gamma$.
\item For each existing vertex, new random edges are added at a rate
      $\lambda$, connecting it to random existing vertices.
\item Each existing edge is deleted at a rate $\mu$.
\end{enumerate}
It is easy to see that the expected order of the graph grows with time
$t$ as $e^{\gamma t}$, and after an initial transient, vertices are
only distinguished by their age, and we are led to consider a
inhomogeneous model with a continuum of vertex types, $\T=[0,\infty[$,
given by vertex age.

%%%%%%%%%%%%%%%%%%%%%%%%%%%%%%%%%%%%%
\subsubsection{Asymptotic properties}

The probability density for ages $x$ is asymptotically given by
\beq
	r(x) = \gamma e^{-\gamma x}.
\eeq
For each pair of vertices, the probability of a connection is
independent of the existence of other connections, and depends on the
age $x$ of the youngest vertex involved, and amounts to, at time $t$,
$p(x) = \frac{2\lambda}{\gamma-\mu} \( e^{(\gamma-\mu)x} - 1 \)
e^{-\gamma t}$.  We obtain
\beq
   c(x,y) = \frac{2\lambda}{\gamma-\mu} \( e^{(\gamma-\mu)\min(x,y)} - 1 \),
\eeq
which seems feasible enough: $c$ is ergodic, continuous and although
$c(x,y)$ is not uniformly bounded, the average $C(x)\equiv\int c(x,y)
r(y) \d y = \frac{2\lambda}{\mu}(1-e^{-\mu x})$ is (at least for
$\mu>0$).  Thus, we are lead to consider the spectrum of the integral
kernel
\beq
	G(x,y) = \frac{2\gamma\lambda}{\mu-\gamma}
	\( 1 - e^{-(\mu-\gamma)\min(x,y)} \) e^{-\gamma y},
\eeq
which is recognized as being proportional to the Green's function
(i.e. a kernel representation of the formal operator inverse)
for a particular differential operator $\LL$ on $\R_+$, given by
\beq
	\LL = -\frac{1}{2\gamma\lambda} e^{\mu x}
	\ddx \(\ddx+\mu-\gamma\),
\eeq
with boundary conditions $f(0)=0$, and $f(x)e^{(\mu-\gamma)x/2}$
growing at most as a power of $x$ as $x\to\infty$.  Criticality
results when the ground state of $\LL$ has eigenvalue 1.

With a {\em finite memory}, $\mu > 0$, the eigenvalue equation for
$\LL$ is a disguised version of Bessel's equation of order
$\gamma/\mu-1$ in the variable $y = \sqrt{8\lambda\gamma} e^{-\mu x/2}
/ \mu$, and criticality results when the first positive zero
$X_{\gamma/\mu-1}$ of $J_{\gamma/\mu-1}$ is given by
$\sqrt{8\lambda\gamma} / \mu$, i.e. for $\lambda = \mu^2
X^2_{\gamma/\mu-1} / 8 \gamma$.

In the special case of {\em infinite memory}, $\mu=0$, the model
reduces to a {\em randomly grown network} \cite{Call01}, and yields
\beq
	\LL = -\frac{1}{2\gamma\lambda} \ddx \(\ddx - \gamma\),
\eeq
with eigenfunctions of the form $e^{\gamma x/2} sin (\omega x)$ with
eigenvalue $\frac{\gamma^2+4\omega^2}{8\gamma\lambda}$, yielding the
ground state value $\frac{\gamma}{8\lambda}$ for $\omega=0$, and
criticality for $\lambda = \gamma/8$.

The above results are all consistent with those obtained in
Refs. \cite{Turo02,Call01} on the phase structure for these models.

%%%%%%%%%%%%%%%%%%%%
\section{Conclusion}

We have investigated a generalization of the classical homogeneous
model of sparse random graphs, obtained by imposing a type structure
on the vertices. This yields a very general class of inhomogeneous
random graph models, and the asymptotic degree distributions are not
restricted to Poissonians, but allow for various types of behaviour,
within certain limitations.  Thus, e.g., power behavior is possible,
while a fixed degree (regular graph) is ruled out.

The models in this class are {\em not} determined by the degree
distribution alone, but contains an infinity of models for each
possible distribution, in contrast to a recently considered class of
models based on a given degree distribution. Interestingly enough, a
relation does exist, since such models are shown to result in a
special case of the present approach.

In other special cases it describes the asymptotic static properties
of certain models of evolving random graphs, such as randomly grown
networks, and dynamical graphs with memory.

Only certain aspects of the approach have been covered in this paper,
and a more detailed analysis, e.g. of the feasibility conditions for
extended type spaces, will be the subject of forthcoming work, as will
investigations on further extensions of the approach e.g. to directed
graphs.

\begin{acknowledgments}
The author wishes to thank Tatyana Turova for stimulating discussions
-- and indeed for an introduction to the subject of random graphs.
This work was in part supported by the Swedish Foundation for
Strategic Research.
\end{acknowledgments}

%%%%%%%%%%%%%%%%%%%%%%%%%%%%%%%
%\bibliographystyle{apsrev}
%\bibliography{RG}

% copied from bbl file:

\end{document}